\documentclass[epj]{webofc}
\usepackage[varg]{txfonts}   
\usepackage{color}
\wocname{EPJ Web of Conferences}
\woctitle{CONF12}
\begin{document}
\selectlanguage{english}
\title{Form factors and differential distributions in rare radiative leptonic B-decays}
%
%

\author{Anastasiia Kozachuk\inst{1,2}\fnsep\thanks{\email{anastasiia.kozachuk@cern.ch}} \and
        Nikolai Nikitin\inst{1,2,3,4}
}

\institute{D.~V.~Skobeltsyn Institute of Nuclear Physics, M.~V.~Lomonosov
Moscow State University, 119991, Moscow, Russia 
\and
           M.~V.~Lomonosov Moscow State University, Faculty of Physics, 119991 Moscow, Russia 
\and
           A.~I.~Alikhanov Institute for Theoretical and Experimental Physics, 117218 Moscow, Russia
\and
           National Research Nuclear University MEPhI, 115409 Moscow, Russia
}

\abstract{
We study rare radiative leptonic decays $B_{(s)}\to e^+e^-\gamma$ and $B_{(s)}\to \mu^+\mu^-\gamma$ within relativistic quark model. In addition to previous analysis we give the estimations of the branching ratios for four values of the minimal photon energy, which correspond to photon selection criteria of the Belle and LHCb detectors. We find out that the branching ratios only slightly change. The highest values corresponding to $E^\gamma_{min}=80 \textrm{MeV}$ are 
${\cal B}(\bar B^0_s\to e^+e^-\gamma)=18.5\times10^{-9}$ and  
${\cal B}(B^0_s\to \mu^+\mu^-\gamma)=11.9\times10^{-9}$. We present the distribution of the forward-backward asymmetry.
}
\maketitle
\section{Introduction}
\label{intro}
At the quark level rare radiative leptonic $B_{(s)}\to\ell^+\ell^-\gamma$ decays correspond to $b\to \{d,s\}$ transitions, which proceed through flavor-changing neutral currents (FCNCs) forbidden at tree level in the Standard Model (SM). These currents are described by penguin and box diagrams containing loops and thus lead to small values of the branching ratios. The typical order is $10^{-8}$ -- $10^{-10}$ (see e.g. \cite{Ali:1996vf}). Possible contributions of new particles to the loops make these decays sensitive to New Physics. Therefore there have been a lot of studies in this sector of flavor physics and as a result at the moment we have several deviations from the SM of the order of $2-4\sigma$ (see discussion in \cite{Glashow:2014iga,Guadagnoli:2015nra,Guadagnoli:2016erb}). We list some of them here:
\begin{itemize}
\item the ratio 
\begin{eqnarray*}
{\cal R}_K\equiv\frac{{\cal B}(B^+\to K^+\mu^+\mu^-)}{{\cal B}(B^+\to K^+e^+e^-)}=0.745^{+0.090}_{-0.074}(\textrm{stat})\pm 0.036(\textrm{syst})
\end{eqnarray*}
in the range $q^2\in[1,6]\,\textrm{GeV}^2$ ($q$ is the invariant mass of the lepton pair) is 25\% lower than the SM prediction at $2.6\sigma$ \cite{Aaij:2014ora, Bobeth:2007dw, Bouchard:2013mia, Hiller:2003js};
\item in an independent measurement, the branching ratio
\begin{eqnarray*}
{\cal B}(B^+\to K^+\mu^+\mu^-)=(1.19\pm 0.03\pm 0.06)\times 10^{-7}
\end{eqnarray*}
is 30\% lower than the SM value at $2\sigma$ \cite{Aaij:2014pli,Aaij:2012vr,Bobeth:2011gi,Bobeth:2011nj,Bobeth:2012vn};
\item same was observed for $B^0_s\to\phi\mu^+\mu^-$, in the range  $q^2\in[1,6]\,\textrm{GeV}^2$ the discrepancy for the branching ratio is more than $3\sigma$ \cite{Aaij:2015esa}.
\end{itemize}
Indeed, more tensions come from angular analysis of $B\to K^*\mu\mu$ performed by LHCb \cite{Aaij:2015oid} and Belle \cite{Abdesselam:2016llu}, and from measurements of the ratios ${\cal R}_{D^{(*)}}={\cal B}(B\to D^{(*)}\tau\nu)/{\cal B}(B\to D^{(*)}\ell\nu)$ \cite{Lees:2012xj,Aaij:2015yra,Huschle:2015rga}. 

Finally, the value of the branching ratio of $B_s\to \mu^+\mu^-$ is 25\% lower than the SM predicts, but only at $1\sigma$. For the branching ratios of $B_s\to \mu^+\mu^-$ and $B_s\to \mu^+\mu^-\gamma$ decays the following relation takes place
\begin{eqnarray}
\frac{{\cal B}(B_s\to \ell^+\ell^-\gamma)}{{\cal B}(B_s\to \ell^+\ell^-)} \,\sim\,\left(\frac{M_{B^0}}{m_{\ell}}\right)^2\frac{\alpha_{em}}{4\pi}, 
\end{eqnarray}
were the squared ratio of masses $(M_{B^0}/m_\ell)^2$ means that the radiative decay $B_s\to \mu^+\mu^-\gamma$ does not have the chirality constraint, $\alpha_{em}$ comes from the photon emission and $4\pi$ in the denominator is the difference between three- and two-particle phase space. For muons one can easily get the estimate $(M_{B^0}/m_\mu)^2\,\sim\,2.5\times 10^3\,\sim\,4\pi/\alpha_{em}$, which means that the branching ratios are approximately equal ${\cal B}(B_s\to \mu^+\mu^-\gamma)\,\sim\,{\cal B}(B_s\to \mu^+\mu^-)$. In fact, ${\cal B}(B_s\to \mu^+\mu^-\gamma)$ is a little bit larger due to additional dynamical effects, such as resonant contributions.

The paper is organized as follows: In Section \ref{sec-1} we discuss the contributions to the decay amplitude $\langle \gamma\ell^+\ell^-|H_{\rm eff}(b\to q \ell^+\ell^-)|B\rangle$. In Section \ref{sec-2} we calculate the transition form-factors via dispersion approach based on constituent quark picture. In Section \ref{sec-3} we give numerical predictions for the branching ratios, differential distributions of the decay rates and forward-backward asymmetry.  
\section{The effective Hamiltonian and the amplitude}
\label{sec-1}
The effective Hamiltonian describing the $b\to q$ ($q=d,s$) weak transition has the form (\cite{Grinstein:1988me, Buras:1994dj})
\begin{eqnarray}
\label{heff}
{\cal H}_{\rm eff}^{b\,\to\, q} = \frac{G_F}{\sqrt{2}} V_{tb} V_{tq}^\ast\, 
\sum_i C_i(\mu) \, O_i(\mu),
\end{eqnarray}
where $G_F$ is the Fermi constant, $C_i$ are the scale-dependent set of Wilson coefficients, and $O_i$ are the basis operators. For $B$ decays the scale parameter $\mu$ is approximately equal to $5$ GeV. The amplitudes of the basis operators between the initial and the final states may be parameterized in terms of Lorentz-invariant form factors. The calculation of these form factors is the main challenge of this work because they contain non-perturbative QCD effects.  
\subsection{Direct emission of the real photon from B-meson valence quarks}
\label{sec-1.1}
The most important part which contains non-perturbative QCD contributions corresponds to the cases, when the real photon is directly emitted from the valence $b$ or $d$ quarks, and the $\ell^+\ell^-$ pair is coupled to the penguin. The effective Hamiltonian in this case takes the form\footnote{Our notations and conventions are: $\gamma^5=i\gamma^0\gamma^1\gamma^2\gamma^3$, $\sigma_{\mu\nu}=\frac{i}{2}[\gamma_{\mu},\gamma_{\nu}]$, $\varepsilon^{0123}=-1$, $\epsilon_{abcd}\equiv\epsilon_{\alpha\beta\mu\nu}a^\alpha b^\beta c^\mu d^\nu$, $e=\sqrt{4\pi\alpha_{\rm em}}$. }  
\begin{eqnarray}
\label{b2qll}
{\cal H}_{\rm eff}^{b\to d\ell^{+}\ell^{-}} = \frac{G_{F}}{\sqrt2}\frac{\alpha_{\rm em}}{2\pi}V_{tb}V^*_{tq}\,\Big[-2im_b\, \frac{C_{7\gamma}(\mu)}{q^2}\cdot\bar d\sigma_{\mu\nu}q^{\nu}\left(1+\gamma_5\right)b\cdot{\bar \ell}\gamma^{\mu}\ell \nonumber\\
+ C_{9V}^{\rm eff}(\mu, q^2)\cdot\bar d \gamma_{\mu}\left (1\, -\,\gamma_5 \right)   b \cdot{\bar \ell}\gamma^{\mu}\ell \, +\, C_{10A}(\mu)\cdot\bar d   \gamma_{\mu}\left (1\, -\,\gamma_5 \right) b \cdot{\bar \ell}\gamma^{\mu}\gamma_{5}\ell \,], 
\end{eqnarray} 
\begin{figure}[h]
\centering
\includegraphics[width=13cm,clip]{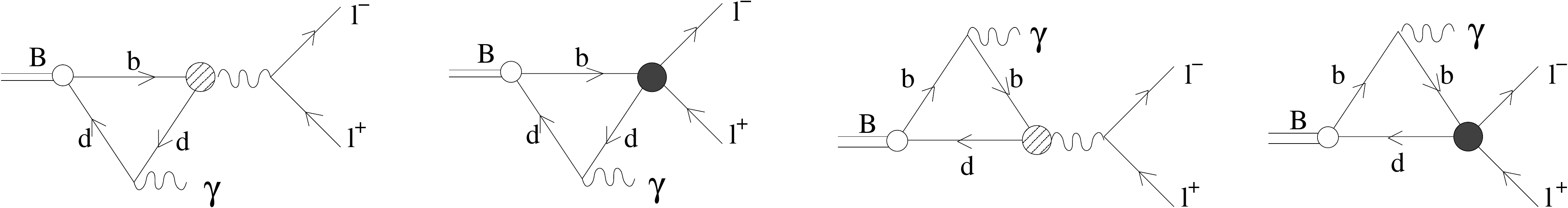}
\caption{Diagrams contributing to 
$B\to\ell^+\ell^-\gamma$ discussed in section \protect\ref{sec-1.1}. Dashed circles denote the $b\to d\gamma$ operator $O_{7\gamma}$. Solid circles denote the $b\to d\ell^+\ell^-$ operators $O_{9V}$ and $O_{10AV}$.}
\label{fig-1}
\end{figure}
and the corresponding diagrams are shown in Fig.~\ref{fig-1}. 
The coefficient $C^{\rm eff}_{9V}(\mu, q^2)$ includes long-distance effects related to $\bar cc$ resonances in the $q^2$-channel \cite{Kruger:1996dt,Melikhov:1998ws,Melikhov:1997wp}. The $B\to \gamma$ transition form factors of the basis operators in (\ref{b2qll}) are defined according to \cite{Kruger:2002gf}
\begin{eqnarray}
\label{real}
\langle
  \gamma (k,\,\epsilon)|\bar d \gamma_\mu\gamma_5 b|B(p) 
\rangle 
&=& i\, e\,\epsilon^*_{\alpha}\, 
\left ( g_{\mu\alpha} \, pk-p_\alpha k_\mu \right )\,\frac{F_A(q^2)}{M_B}, 
\nonumber
\\
\langle
  \gamma(k,\,\epsilon)|\bar d\gamma_\mu b|B(p)
\rangle
&=& 
e\,\epsilon^*_{\alpha}\,\epsilon_{\mu\alpha\xi\eta} p_{\xi}k_{\eta}\, 
\frac{F_V(q^2)}{M_B},   
\\
\langle
  \gamma(k,\,\epsilon)|\bar d \sigma_{\mu\nu}\gamma_5 b|B(p) 
\rangle\, (p-k)^{\nu}
&=& 
e\,\epsilon^*_{\alpha}\,\left[ g_{\mu\alpha}\,pk- p_{\alpha}k_{\mu}\right ]\, 
F_{TA}(q^2, 0), 
\nonumber
\\
\langle
\gamma(k,\,\epsilon)|\bar d \sigma_{\mu\nu} b|B(p) 
\rangle\, (p-k)^{\nu}
&=& 
i\, e\,\epsilon^*_{\alpha}\epsilon_{\mu\alpha\xi\eta}p_{\xi}k_{\eta}\, 
F_{TV}(q^2, 0).
\nonumber 
\end{eqnarray}
The penguin form factors $F_{TV,TA}(q_1^2,q_2^2)$ are defined as functions of two variables: $q_1$ is the momentum of the photon emitted from the penguin, and $q_2$ is the momentum of the photon emitted from the valence quark of the $B$ meson. We calculate the form factors in the framework of the dispersion approach based on constituent quark picture, the details are presented in Section \ref{sec-2}. 
\subsection{Direct emission of the virtual photon from B-meson valence quarks}
\label{sec-1.2}
Another process contributing to the amplitude is that with the real photon emitted from the penguin, whereas one of the valence quarks directly emits the virtual photon which then goes into the final $\ell^+\ell^-$ pair. 
\begin{figure}[h]
\centering
\includegraphics[width=7cm,clip]{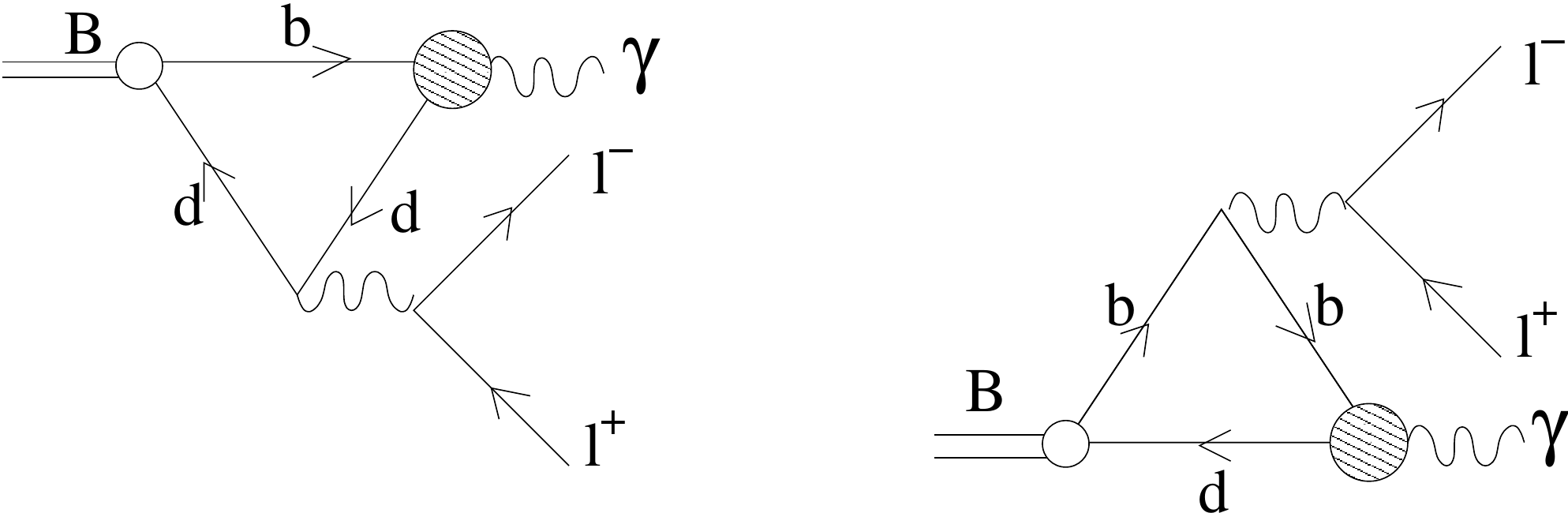}
\caption{Diagrams contributing to $B\to\ell^+\ell^-\gamma$ discussed in section \protect\ref{sec-1.2}. Dashed circles denote the $b\to d\gamma$ operator $O_{7\gamma}$.}
\label{fig-2} 
\end{figure}
This process is described by the diagrams of Fig.~\ref{fig-2}. 
The corresponding amplitude has the same structure as the $C_{7\gamma}$ part of the amplitude in \ref{sec-1.1} with $F_{TA,TV}(q^2,0)$ replaced by $F_{TA,TV}(0,q^2)$. The form factors $F_{TA,TV}(0,q^2)$ at the necessary timelike momentum transfers are not known. The difficulty is connected with appearance of neutral light vector meson resonances, $\rho^0$ and $\omega$ for $B$-decays and $\phi$ for $B_s$-decays, in the physical $B\to \gamma \ell^+\ell^- $ decay region. We calculate the form factors $F_{TA,TV}(0,q^2)$ for $q^2>0$ with the use of gauge-invariant version \cite{Melikhov:2001ew,Melikhov:2003hs} of the vector meson dominance \cite{Sakurai:1960ju,GellMann:1961tg,Gounaris:1968mw}
\begin{eqnarray}
\label{zamena1a}
F_{TV,TA}(0, q^2) = F_{TV,TA}(0, 0)\, -\,\sum_V\,2\,f_V g^{B\to V}_+(0)\frac{q^2/M_V}{q^2\, -\, M^2_V\, +\, iM_V\Gamma_V},
\end{eqnarray}
where $M_V$ and $\Gamma_V$ are the mass and the width of the vector meson resonance, $g^{B\to V}_+(0)$ are the $B\to V$ transition form factors, defined according to the relations 
\begin{eqnarray*}
\langle V(q, \varepsilon)|\bar d\sigma_{\mu\nu} b|
B(p)\rangle
\, =\, i\varepsilon^{*\alpha}\,\epsilon_{\mu\nu\beta\gamma}
\left[ 
g^{B\to V}_+(k^2)g_{\alpha\beta}(p+q)^{\gamma} + g^{B\to V}_-(k^2)g_{\alpha\beta}k^{\gamma} + 
g^{B\to V}_0(k^2)p_{\alpha}p^{\beta}q^{\gamma}
\right] 
\end{eqnarray*}
and calculated in \cite{Melikhov:1995xz,Melikhov:1997qk} via relativistic quark model. The leptonic decay constant of a vector meson is given by
\begin{eqnarray}
\langle 0|\bar d \gamma_\mu d|V(\varepsilon, p)\rangle=\varepsilon_\mu M_V f_V.  
\end{eqnarray}
\subsection{Bremsstrahlung}
\label{sec-1.3}
\begin{figure}[h]
\centering
\includegraphics[width=3cm,clip]{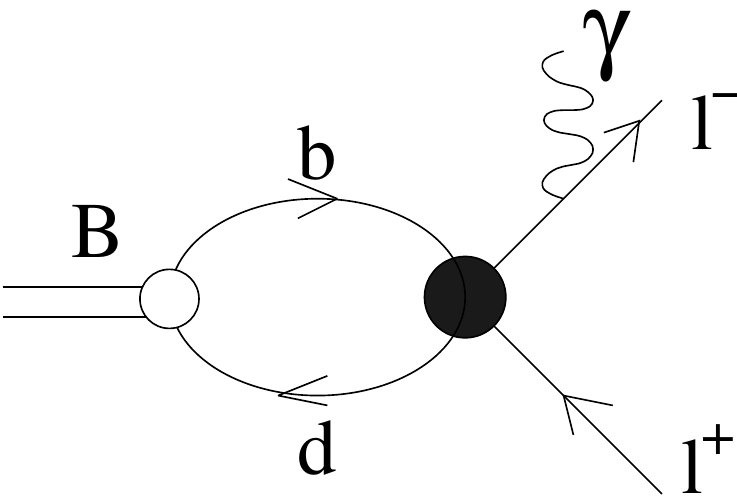}
\caption{Diagrams describing photon bremsstrahlung. Solid circles denote the operator $O_{10A}$.}
\label{fig-3} 
\end{figure}
Fig.~\ref{fig-3} represents diagrams describing bremsstrahlung. The corresponding contribution to the 
$B\to \ell^+\ell^-\gamma$ amplitude  reads 
\begin{eqnarray}
\label{bremsstrahlung}
A_\mu^{\rm Brems}=-i\, e\,\frac{G_F}{\sqrt{2}}\,\frac{\alpha_{\rm em}}{2\pi}\, V^*_{td}V_{tb}\, 
\frac{f_{B_q}}{M_B}\, 2\hat m_{\ell}\, C_{10A}(\mu)\, 
\bar\ell (p_2)
\left [
\frac{(\gamma\epsilon^*)\,(\gamma p)}{\hat t-\hat m^2_{\ell}}\, -\, 
\frac{(\gamma p)\,(\gamma\epsilon^*)}{\hat u-\hat m^2_{\ell}}
\right ]
\gamma_5\,\ell (-p_1),
\end{eqnarray}
$f_{B_q}>0$. 
\subsection{Weak annihilation contribution}
\label{sec-1.4}
\begin{figure}[h]
\centering
\includegraphics[width=5cm,clip]{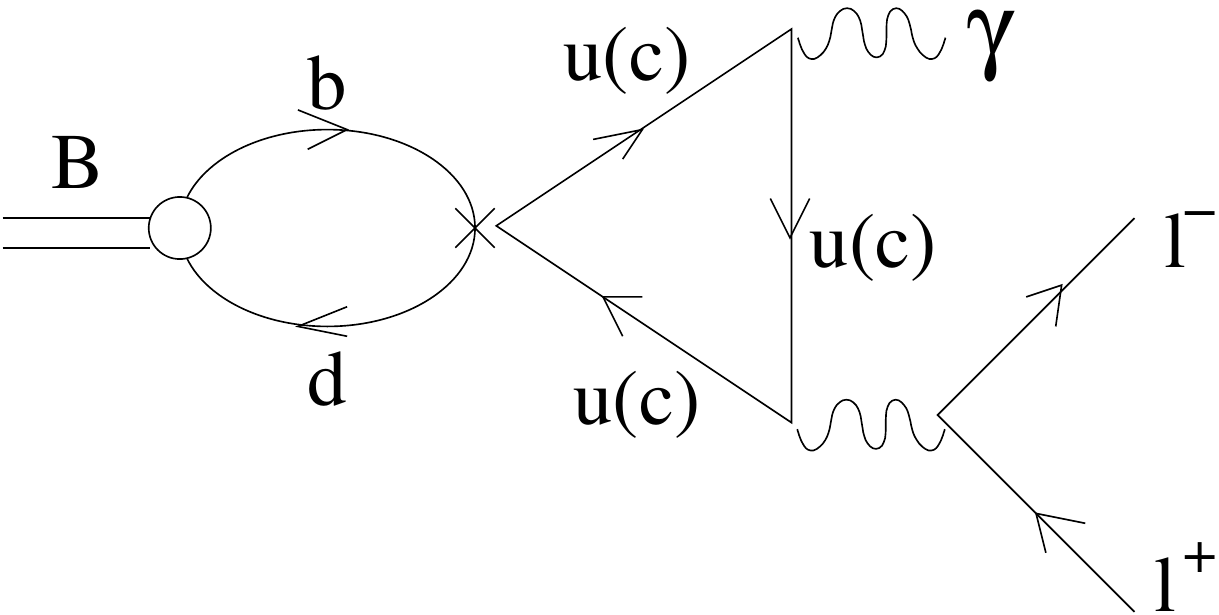}
\caption{Weak annihilation diagrams contributing to the process.} 
\label{fig-4}   
\end{figure}
The weak annihilation contribution is given by triangle diagrams of Fig.~\ref{fig-4}. One should take into account $u$ and $c$ quarks in the loop. 
The vertex describing the $\bar bd\to \bar QQ$ 
transition ($Q=u,c$) reads
\begin{eqnarray}
\label{heffwa}
H_{\rm eff}^{B\to\bar QQ} = -\,\frac{G_F}{\sqrt{2}}\, a_1\,V_{Qb}V^*_{Qd}
\,\bar d\gamma_{\mu}(1 -\gamma_5)b 
\,\bar Q\gamma_{\mu}(1 -\gamma_5)Q, 
\end{eqnarray}
with $a_1\, =\, C_1\, +\, C_2/N_c$, $N_c$ number of colors \cite{Neubert:1997uc}. 
\section{Transition form factors}
\label{sec-2} 
We calculate the transition form factors in the framework of relativistic quark model, which is a dispersion approach based on constituent quark picture \cite{Melikhov:1995xz,Melikhov:1997qk}. All hadron observables are given by dispersion representations in terms of hadronic relativistic wave functions and spectral densities of corresponding feynman diagrams with constituent quarks in the loops. For the wave functions we use a Gaussian parametrization $\phi(s)=A(s,\beta)e^{-k^2(s)/(2\beta^2)}$. The simplest relation can be obtained for a pseudoscalar or vector meson decay constant
\begin{eqnarray} \label{eq:fM}
f_M=\int ds\phi_M(s)\rho(s),
\end{eqnarray}
were $\phi(s)$ is the meson relativistic wave function and $\rho(s)$ is the spectral density. The latter is obtained as a direct result of feynman rules for the corresponding feynman diagram. 
\begin{figure}[h]
\centering
\includegraphics[width=3cm,clip]{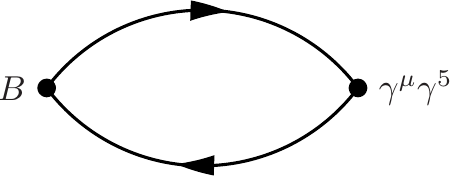}
\caption{Feynman diagram corresponding to dispersion representation of B-meson decay constant.}
\label{fig-5}       
\end{figure}
The example of the diagram for a B-meson decay constant is given in Fig.~\ref{fig-5}.
In this work we consider meson-to-photon transitions; the corresponding form factors $F_{V,A,TV,TA}$ may be 
obtained in the form of the spectral representation 
\begin{eqnarray}
F(q_1,q_2)=\int ds\phi(s)\frac{ds'\Delta(s,s',q_2^2)}{s'-q_1^2},
\end{eqnarray} 
were $q_1$ and $q_2$ are momenta of the emitted photons. The form factors $F_{V,A}$ were calculated 
in \cite{Kozachuk:2015kos}. We now perform the calculation of the form factors $F_{TV,TA}$. 
Each of these form factors contains two contributions corresponding to the cases when the photon is emitted from $b$ or $d(s)$ quarks of the B-meson: 
\begin{eqnarray}
\label{FTV}
F_{TV}\,=\,Q_dF^{(1)}_{TV}(m_d,m_b)+Q_bF^{(1)}_{TV}(m_b,m_d),\nonumber \\
F_{TA}\,=\,Q_dF^{(1)}_{TA}(m_d,m_b)+Q_bF^{(1)}_{TA}(m_b,m_d).  
\end{eqnarray}
The spectral representations of the form factors in (\ref{FTV}) have the form 
\begin{eqnarray}
F_{TV}^{(1)}(s) &=& -\int\limits_{(m_1+m_2)^2}^\infty ds g_2(s, m_1, m_2)-\frac{M_B^2+q^2}{M_B^2-q^2}\int\limits_{(m_1+m_2)^2}^\infty ds g_1(s, m_1, m_2), \nonumber\\				
F_{TA}^{(1)}(s) &=& -\int\limits_{(m_1+m_2)^2}^\infty ds g_2(s, m_1, m_2) - \int\limits_{(m_1+m_2)^2}^\infty ds g_1(s, m_1, m_2),
\end{eqnarray} 
where $m_1$ is the mass of the quark, which emits the photon, $m_2$ is the mass of the spectator, and
\begin{eqnarray*}
g_1(s, m_1, m_2) = \phi_B(s,m_1,m_2) \frac{M_B^2 - q^2}{(s-q^2)^2}\,\Bigg(\,\frac{s + m_1^2 - m_2^2}{2s}\sqrt{\lambda(s, m_1, m_2)} - \\
- m_1^2 \log{\frac{s + m_1^2 - m_2^2 + \sqrt{\lambda(s, m_1, m_2)}}{s + m_1^2 - m_2^2 - \sqrt{\lambda(s, m_1, m_2)}}}\,\Bigg)\,, \\
\end{eqnarray*}
\begin{eqnarray}
g_2(s, m_1, m_2) = \phi_B(s,m_1,m_2) \frac{1}{s - q^2}\,\Bigg(\,\sqrt{\lambda(s, m_1, m_2)} - \nonumber\\
 m_1(m_2-m_1) \log{\frac{s + m_1^2 - m_2^2 + \sqrt{\lambda(s, m_1, m_2)}}{s + m_1^2 - m_2^2 - \sqrt{\lambda(s, m_1, m_2)}}}\,\Bigg)\,.
\end{eqnarray}
The model contains only a few parameters such as the constituent quark masses and the parameter of the wave function $\beta$. 
These parameters were fixed in \cite{Kozachuk:2015kos} using relations (\ref{eq:fM}) for meson decay constants 
so that our results reproduce the predictions from QCD sum rules and lattice QCD. 
\section{Numerical results}
\label{sec-3}
\subsection{Branching ratios}
\label{sec-3.1}
\begin{figure}[h]
\begin{center}
\begin{tabular}{cc}
\mbox{\centering
\includegraphics[width=6cm,clip]{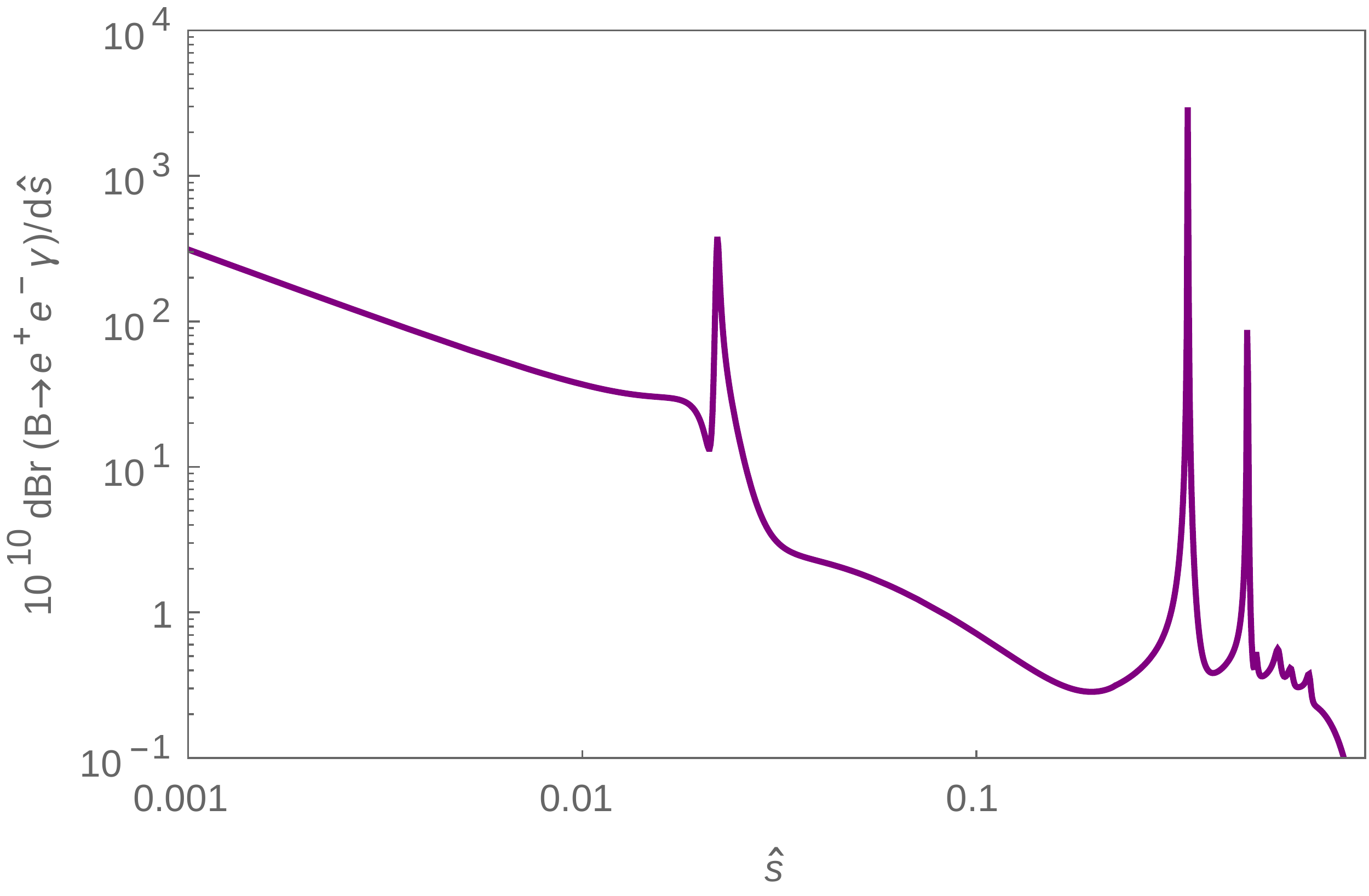}} &
\mbox{\centering
\includegraphics[width=6cm,clip]{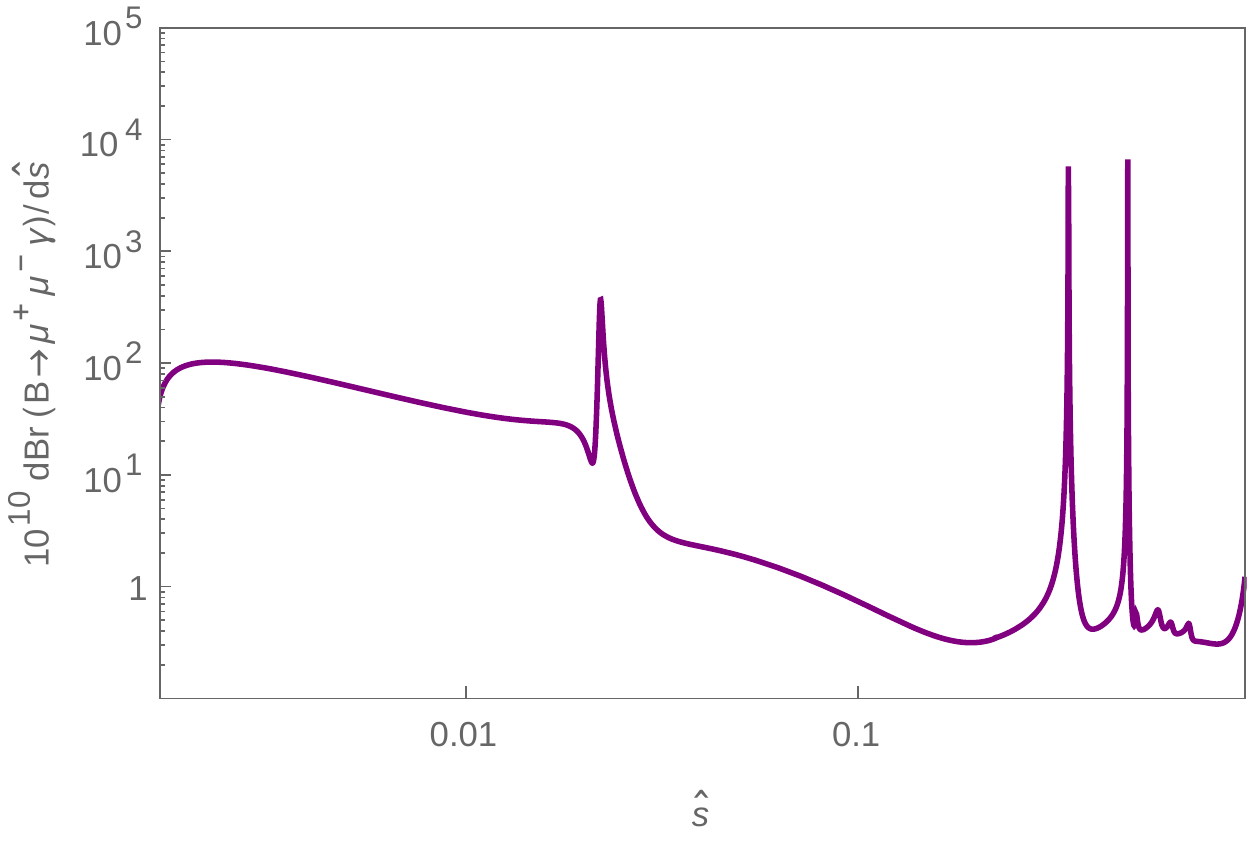}} 
\end{tabular}
\caption{Differential branching fractions for $B\to e^+e^-\gamma$ (left) and $B\to \mu^+\mu^-\gamma$ (right) decays.}
\label{fig-6} 
\end{center}
\end{figure}
\begin{figure}[h]
\begin{center}
\begin{tabular}{cc}
\mbox{\centering
\includegraphics[width=6cm,clip]{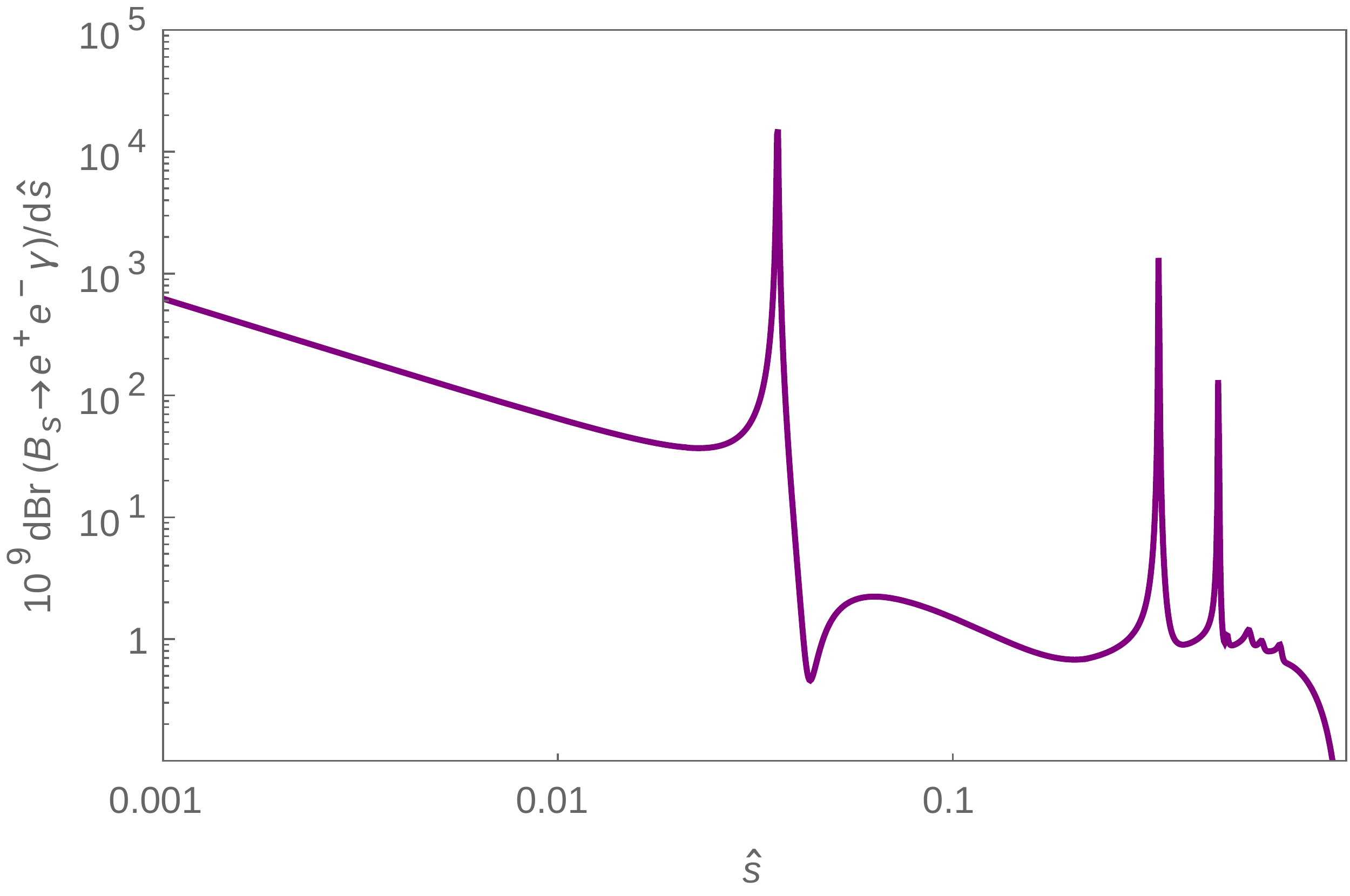}} &
\mbox{\centering
\includegraphics[width=6cm,clip]{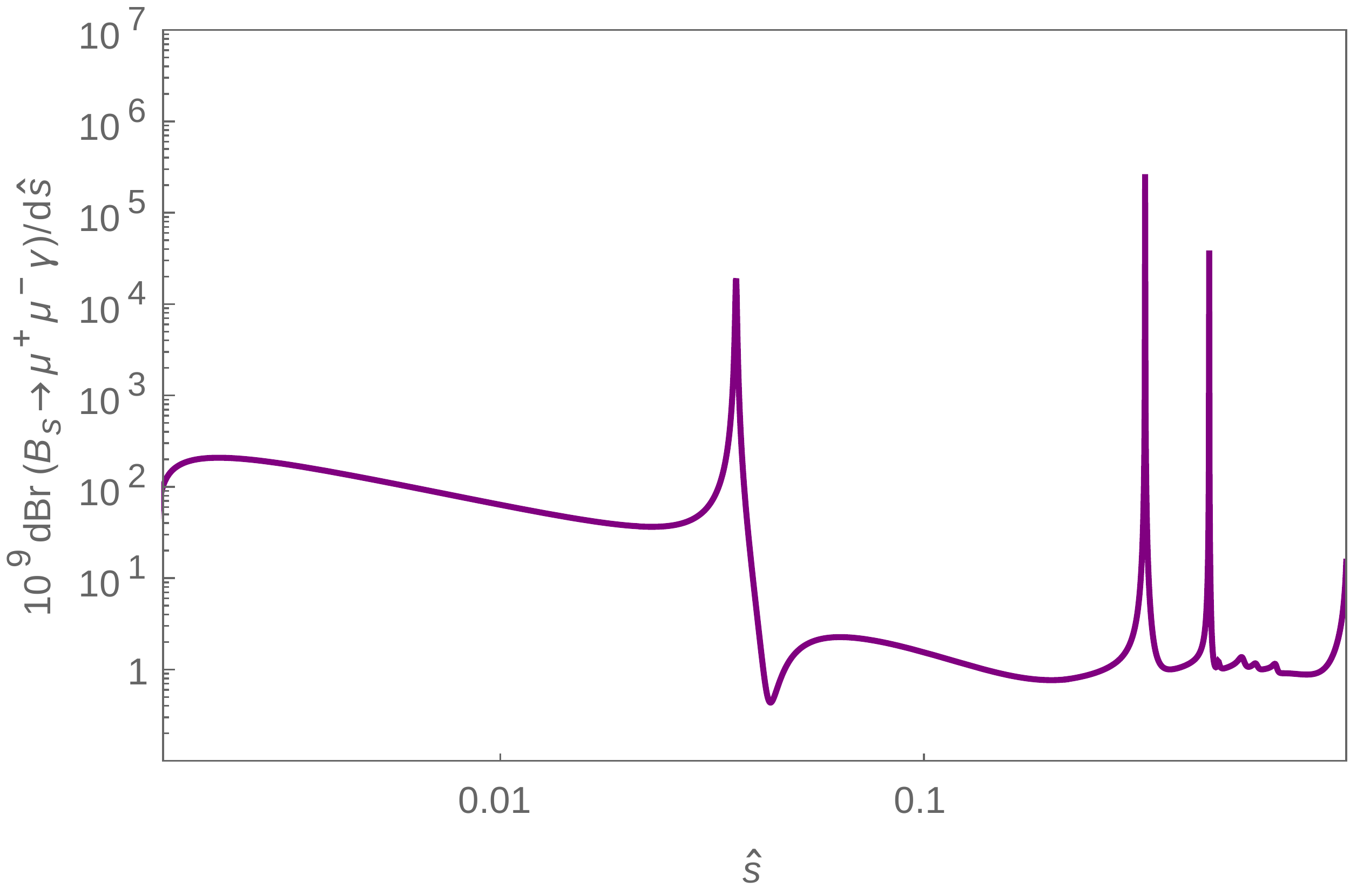}} 
\end{tabular}
\caption{Differential branching fractions for $B_s\to e^+e^-\gamma$ (left) and $B_s\to \mu^+\mu^-\gamma$ (right) decays.}
\label{fig-7} 
\end{center}
\end{figure}
For numerical estimates we use the following values of Wilson coefficients at $\mu=5$ GeV: \\
$C_1(5\, GeV)=\,0.235$, $C_2(5\, GeV)=\, -1.1$, $C_{7\gamma}(5\, GeV)=\, 0.308$, $C_{10A}(5\, GeV)=4.63$ \cite{Buras:1994dj}. The $C^{eff}_{9V}(\mu,s)$ evolution including cc-resonances is taken from \cite{Kruger:1996dt,Melikhov:1998ws,Melikhov:1997wp}.
We obtained several distributions of the differential branching ratios, they are shown in Fig.~\ref{fig-6} and \ref{fig-7}. 
\begin{table}[h]
\centering
\caption{Our results for the branching ratios of $B_{(s)}\to l^+l^-\gamma$ decays for different photon energy cuts}
\label{tab-1}
\begin{tabular}{l|llll}
\hline
$E^\gamma_{min}$                    & 80 MeV & 100 MeV & 500 MeV & 1000 MeV \\
\hline
\protect\( Br\left ( B\to e^+e^-\gamma\right )\,\times\, 10^{10}\protect\) &
                           4.6 &
                           4.6 &
                           4.6 & 
                           4.6 \\
\protect\( Br\left ( B\to \mu^+\mu^-\gamma\right )\,\times\, 10^{10}\protect\) &
                           1.5 &
                           1.5 &
                           1.5 & 
                           1.4 \\
\protect\( Br\left ( B_s\to e^+e^-\gamma\right )\,\times\, 10^{9}\protect\) &
                           18.5 &
                           18.5 &
                           18.4 &
                           18.3 \\
\protect\( Br\left ( B_s\to \mu^+\mu^-\gamma\right )\,\times\, 10^{9}\protect\) &
                           11.9 &
                           11.8 &
                           11.6 &
                           11.5 \\
\hline
\end{tabular}
\end{table}
In Table~\ref{tab-1} we present the results for different values of the minimal photon energy $E^\gamma_{min}$. Numerical values of $E^\gamma_{min}$ are given in the $B$-meson rest frame and are different in others because energy is not Lorentz-invariant. The values in the range $E^\gamma_{min}\in[100,500]\,\textrm{MeV}$ correspond to different photon selection criteria at the Belle II detector \cite{Aushev:2010bq}, while the interval $E^\gamma_{min}\in[500,1000]\,\textrm{MeV}$ is relevant for those at the LHCb detector \cite{LHCB:2000ab,Voss:2009zz}. We only take typical values as in LHCb reference frame B-mesons have a sufficiently wide energy distribution. However, from Table~\ref{tab-1} it is clear that the branching ratios only slightly depend on the particular choice of $E^\gamma_{min}$, and the related error is smaller then that of the model. We also use the value $E^\gamma_{min}=80\,\textrm{MeV}$ to compare the results with those of the previous work \cite{Melikhov:2004mk}.
\subsection{Forward-backward asymmetry}
\label{sec-3.2}
\begin{figure}[h]
\begin{center}
\begin{tabular}{cc}
\mbox{\centering
\includegraphics[width=6cm,clip]{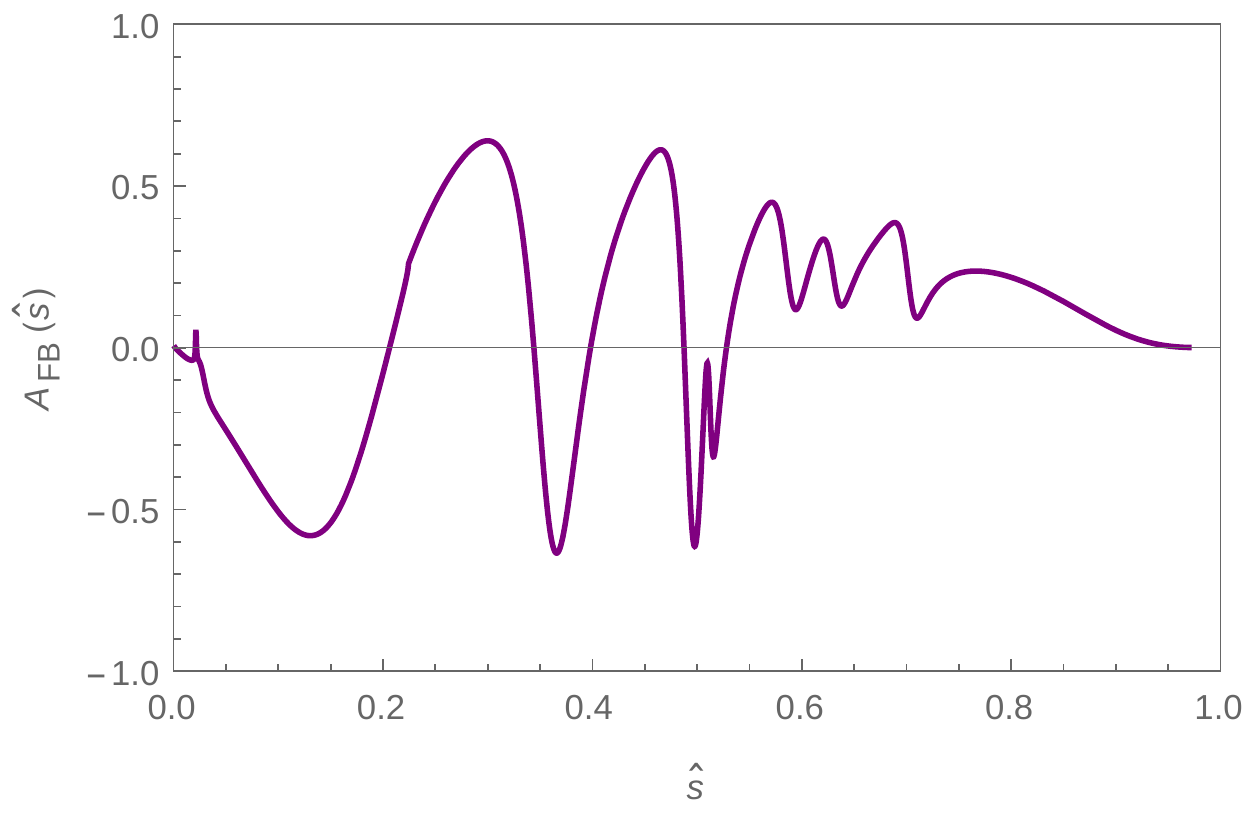}} &
\mbox{\centering
\includegraphics[width=6cm,clip]{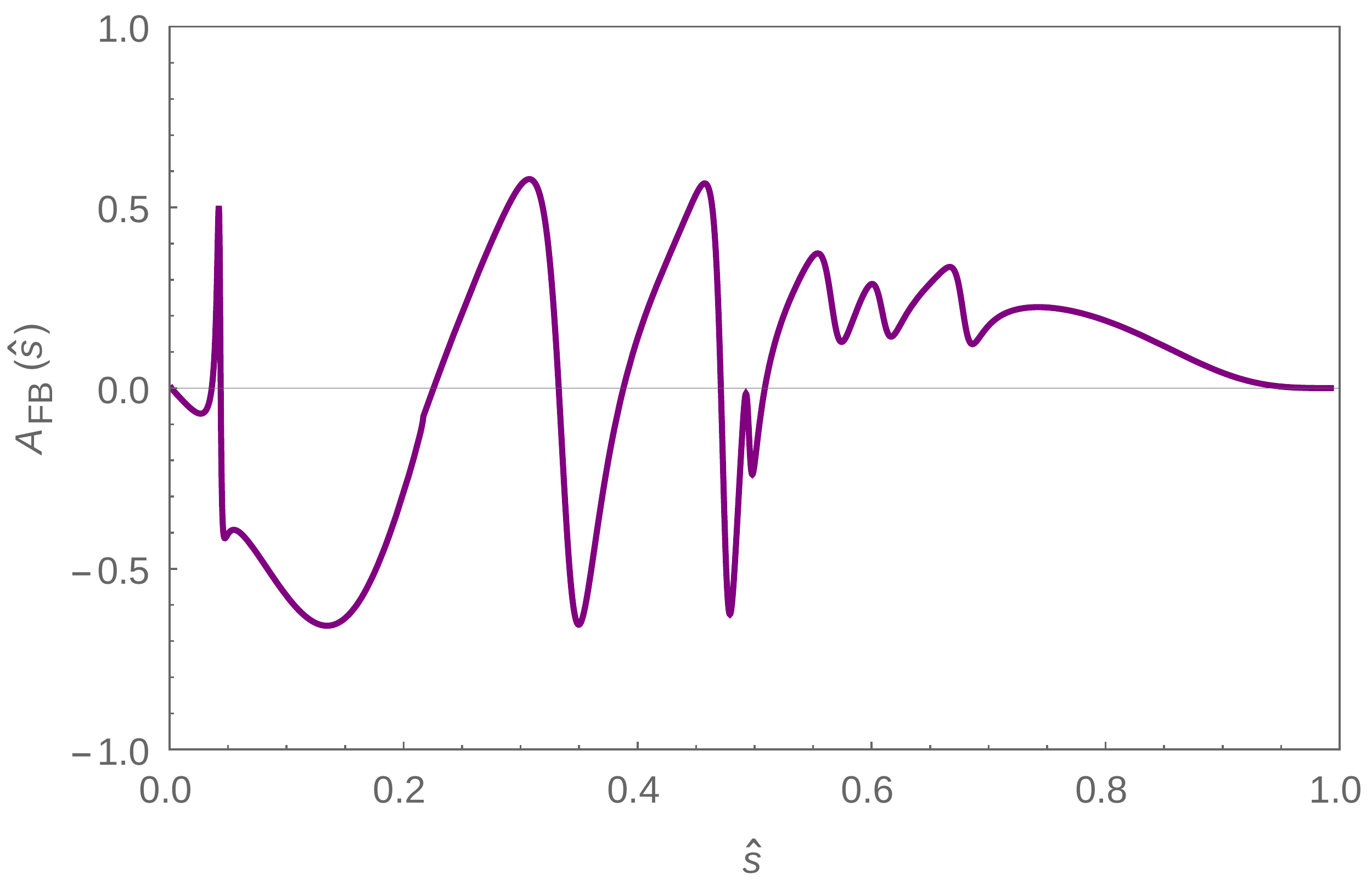}} 
\end{tabular}
\caption{Forward-backward asymmetry for $B\to \mu^+\mu^-\gamma$ (left) and $B_s\to \mu^+\mu^-\gamma$ (right) decays.}
\label{fig-8} 
\end{center}
\end{figure}
We obtained distribution of the forward-backward asymmetry, defined by the relation
\begin{eqnarray}
A_{FB}(\hat{s})\,=\,\frac{\int\limits_0^1d\cos\theta \, \frac{d^2\Gamma(B_{(s)}\to\ell^+\ell^-\gamma)}{d\hat{s} \, d\cos\theta}-\int\limits_{-1}^0d\cos\theta \, \frac{d^2\Gamma(B_{(s)}\to\ell^+\ell^-\gamma)}{d\hat{s} \, d\cos\theta}}{\frac{d\Gamma(B_{(s)}\to\ell^+\ell^-\gamma)}{d\hat{s}}},
\end{eqnarray}
where $\hat{s}\,=\,q^2/M_B^2$,\, $\theta$ is the angle between $\vec{p}$ and $\vec{p_2}$. The distribution is presented in Fig.~\ref{fig-8}.
The measurement of the forward-backward asymmetry seems to be a hard if not impossible task, because the final state $\ell^+\ell^-\gamma$ does not carry any information about the flavor of the decaying B-meson. In addition, the signs of the asymmetries corresponding to $B$ and $\bar B$ mesons are opposite. In the absence of flavor cut the total asymmetry equals zero aside from CP-violating effects. It appears that such selection is impossible at LHCb, but at Belle II one can use the fact that neutral B-mesons are produced in an entangled state. Thus, if one of the B-mesons decays to a state with a certain flavor, the other B-meson which decays to $\ell^+\ell^-\gamma$ has the opposite flavor. Now, if the interval between the decays is less than half of the of oscillation period, one can claim that the flavor of the second B-meson is known with sufficient probability. For each selection procedure one can also account for the oscillations contribution and therefore improve the prediction accuracy.
A method for such calculations was developed in \cite{Balakireva:2009kn}.

The decay rates and the forward-backward asymmetry were previously calculated in several works \cite{Geng:2000fs, Dincer:2001hu, Kruger:2002gf, Melikhov:2004mk, Wang:2013rfa, Balakireva:2009kn}. In \cite{Geng:2000fs} and \cite{Dincer:2001hu} not all the contributions were taken into account, and in \cite{Geng:2000fs,Dincer:2001hu, Kruger:2002gf, Melikhov:2004mk} the transition form factors were estimated from symmetry considerations coming from LEET. 
We made direct calculation of the form factors in the framework of the relativistic quark model. 
Our results agree nicely with \cite{Dincer:2001hu}. 
The results \cite{Geng:2000fs,Wang:2013rfa} are based on not fully consistent models for the form factors and therefore do not seem to us convincing. 
\section{Conclusions}
\label{concl}
We obtained predictions for the differential distributions and the branching ratios for the $B_{(s)}\to e^+e^-\gamma$ and $B_{(s)}\to \mu^+\mu^-\gamma$ decays
taking into account the following contributions to the amplitude of the process: photon emission from the $d(s)$ and the $b$ valence quarks of the $B$-meson, 
weak annihilation, and bremsstrahlung. The corresponding form factors were calculated in the framework of the relativistic quark model. We represented the numerical estimates for several values of the minimal photon energy corresponding to photon selection criteria of the Belle and LHCb detectors and found out that the branching ratios almost do not change with the photon energy cut.
\section{Acknowledgements}
\label{sec:acknownlegements}
The work was supported by grant 16-12-10280 of the Russian Science Foundation. The authors are thankful to Dmitri Melikhov for useful discussions.

\end{document}